\documentclass[reprint, amsmath, amssymb, aps, pra 
]{revtex4-2}

\usepackage{graphicx}
\usepackage{dcolumn}
\usepackage{bm}
\usepackage{hyperref}
\usepackage[export]{adjustbox}
\usepackage{xcolor}
\usepackage[toc,page,titletoc]{appendix}

\usepackage{comment}
\usepackage{tensor}

\begin{document}

\title{Optical probes of coherence in two dimensional Bose gases of polaritons}

\author{Joseph Jachinowski}
\email{jachinowski@uchicago.edu}
\affiliation{James Franck Institute and the Department of Physics, University of Chicago, Chicago, IL 60637, USA}
\author{Hassan Alnatah}
\affiliation{Department of Physics, University of Pittsburgh, 3941 O’Hara Street, Pittsburgh, Pennsylvania 15218, USA}
\author{David W. Snoke}
\affiliation{Department of Physics, University of Pittsburgh, 3941 O’Hara Street, Pittsburgh, Pennsylvania 15218, USA}
\author{Peter B. Littlewood}%
\affiliation{James Franck Institute and the Department of Physics, University of Chicago, Chicago, IL 60637, USA}
\affiliation{School of Physics and Astronomy, University of St Andrews, St Andrews KY16 9SS, United Kingdom }

\date{\today}

\begin{abstract}
Due to their photonic components, exciton-polariton systems provide a convenient platform to study the coherence properties of weakly-interacting Bose gases. In particular, optical interferometry enables the measurement of the first-order coherence function which provides information about the intrinsic correlations of the system. In this paper, we derive a universal curve for the coherent fraction of a noninteracting, equilibrium, homogeneous, two-dimensional Bose gas, with density expressed in units of the observation area, and compare to recent experimental results. Although there is a sharp transition from normal to superfluid phases in the thermodynamic limit, the coherent fraction of the gas varies continuously across this transition due to the finite system size. We find that the theory agrees nearly perfectly with the experimental data in the low-density limit with no free parameters other than the effective temperature, highly constrained by the measurements. At higher density, the experiments are consistent with standard weakly-interacting Bose gas theory. By having a theory that treats both the optical diffraction and Bose coherence, we can clearly see the effect of the quantum statistics on the coherence.
\end{abstract}

                              
\maketitle


Exciton-polaritons (EP) are bosonic matter-light quasiparticles formed by coupling excitons, bound electron-hole pairs in solids, to photons confined in optical cavities \cite{deng_exciton-polariton_2010, timofeev_exciton_2012}. While the excitonic component endows the EPs with interactions, the photonic component keeps the effective mass very small and the particles delocalized, driving the Bose-Einstein condensate (BEC) transition from cryogenic temperatures up towards room temperature \cite{alnatah_boseeinstein_2025, alnatah_strong_2025}.
Various experiments have investigated the signatures of BEC in EP systems \cite{deng_condensation_2002, kasprzak_boseeinstein_2006, deng_quantum_2006, balili_bose-einstein_2007};
besides BEC, EPs systems possess a wide range of accessible phases \cite{littlewood_models_2004} due their multicomponent driven-dissipative nature. Such phases can be probed by measuring the first-order coherence function, and experiments have confirmed the buildup of (quasi)-long-range-order near a critical pump power \cite{deng_spatial_2007} \footnote{Analogous experiments have been performed in photonic gases \cite{damm_first-order_2017}.} and the spontaneous formation of quantized vortices \cite{lagoudakis_quantized_2008} related to superfluidity.

While nonequilibrium pumping and decay processes complicate the system, it is possible to prepare EP samples which are well-described by the equilibrium Bose-Einstein distribution \cite{sun_bose-einstein_2017, alnatah_coherence_2024}. The equilibrium limit provides a baseline for the theory of two-dimensional condensates more broadly. Outside of this regime, other experiments have explored the phenomena of multimode condensation \cite{marelic_spatiotemporal_2016, alnatah_critical_2024}, the Berezinskii-Kosterlitz-Thouless (BKT) phase \cite{berezinskii_destruction_1971, kosterlitz_ordering_1973} in a nonequilibrium setting \cite{roumpos_power-law_2012}, and its interplay with Kardar-Parisi-Zhang (KPZ) physics \cite{kardar_dynamic_1986, diessel_emergent_2022, fontaine_kardarparisizhang_2022}.

In this paper, we study the build up of coherence in EP systems as the density approaches the critical density. Our analysis is based on the first-order coherence function for a noninteracting, equilibrium Bose gas, which we use to define the condensate fraction and the related coherent fraction, directly measurable in interferometry experiments. 
Our principal result is a universal curve for the coherent fraction of a finite-sized system that depends only on the density, expressed in units of the aperture area, the mass of the particles, and their temperature. In the zero-density limit, the coherent fraction approaches a small but nonzero constant given by a simple analytic expression; in the high-density limit, it saturates at one, as shown in Fig.~\ref{fig:coherent fraction no diffraction}.
In addition to explaining the universal behavior observed in the equilibrium EP experiments, our results are broadly applicable to interference experiments with partially coherent sources.

We compare our theory results to recent experimental data from Ref.~\cite{alnatah_coherence_2024} in Fig.~\ref{fig: theory experiment comparison}. We find good agreement in the normal, non-superfluid state using no free parameters. Additionally, we argue that the identified power-law behavior of the coherent fraction applies approximately over a limited range of densities. Then, precisely when the noninteracting theory deviates from the experimental results, the textbook theory for a weakly-interacting superfluid begins to accurately model the data, with a value for the interaction strength that is in the range of experimentally measured values \cite{alnatah_coherence_2024, snoke_reanalysis_2023}. This crossover is captured by the Gross-Pitaevskii equation, simulated in Ref.~\cite{alnatah_coherence_2024}, but the analysis does not yield analytic results connecting to the noninteracting limit. 

\section{Theory of coherence in Bose gases} \label{sec: theory of Bose gases}

The (equal-time) first-order coherence function of a Bose gas $G^{(1)} = \left\langle \Psi^\dagger \left( \bm{r} \right) \Psi \left( \bm{r}' \right) \right\rangle$ can be expressed in two dimensions as
\begin{equation} \label{eq: coherence function basic definition}
    G^{(1)} \left( \bm{r}, \bm{r}' \right) = \int_{\mathbb{R}^2} \frac{d^2 k_\parallel}{\left(2 \pi \right)^2} \; e^{i \bm{k}_\parallel \cdot \left( \bm{r} - \bm{r}' \right)} \left\langle a_{\bm{k}_\parallel}^\dagger a_{\bm{k}_\parallel} \right\rangle
\end{equation}
assuming translation invariance. Here, $\Psi \left( \bm{r} \right)$ is the field operator at position $\bm{r}$, and $a_{\bm{k}_\parallel}$ is the corresponding annihilation operator for momentum mode $\bm{k}_\parallel$ calculated via Fourier transform. In the case of EPs (expanded upon in Supplemental Material (SM) Sec.~\ref{sec: background on coherence}), $\bm{k}_\parallel$ is the momentum in the two dimensional plane perpendicular to the (micro)cavity axis. Assuming an equilibrium Bose gas in the grand canonical ensemble, we use the Bose-Einstein momentum distribution $\tilde{n} \left( \bm{k}_\parallel \right) = \left\langle a_{\bm{k}_\parallel}^\dagger a_{\bm{k}_\parallel} \right\rangle = \left( z^{-1} e^{\lambda_T^2 k_\parallel^2 / 4 \pi} - 1 \right)^{-1}$. The relevant parameters are the fugacity $z = e^{\mu / k_B T}$ expressed in terms of the chemical potential $\mu$, the temperature $T$, and Boltzmann's constant $k_B$, and the thermal de Broglie wavelength $\lambda_T = \sqrt{ 2 \pi \hbar^2 / m k_B T}$ where $m$ is the particle mass, and $\hbar$ is the reduced Planck constant.

\begin{figure}[t]
    \centering
    \includegraphics[width=0.98\linewidth]{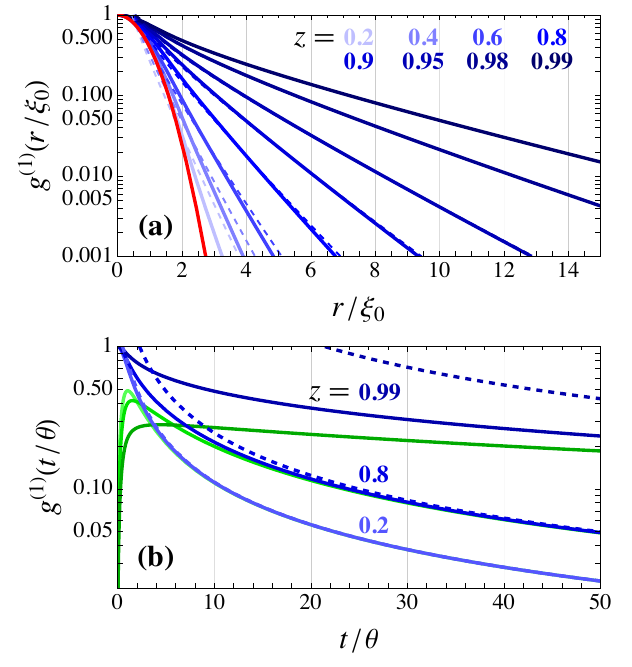}
    \caption{Correlations of a noninteracting, homogeneous Bose gas in two dimensions. (a) Exact (solid blue, Eq.~\ref{eq: coherence function}) and asymptotic (dashed blue, SM Sec.~\ref{sec: background on coherence}) equal-time normalized first-order coherence function $g^{(1)}$ as a function of the normalized position-separation $r / \xi_0$ where $\xi_0 = \lambda_T / \sqrt{\pi}$ is the classical coherence length. (Solid red) Short-range Gaussian approximation of the exact $z = 0.2$ first-order coherence function. (b) Exact (solid blue, SM Sec.~\ref{sec: background on coherence}) and asymptotic (dashed blue, Eq.~\ref{eq: coherence function time}) equal-position normalized first-order coherence function magnitude $\left| g^{(1)} \right|$ as a function of the normalized time-separation $t / \theta$ where $\theta = \hbar / k_B T$. (Solid green) Imaginary part of the exact equal-position normalized first-order coherence function $- \text{Im} \; g^{(1)}$ which dominates at long time-separations.
    }
    \label{fig: correlations}
\end{figure}

The first-order coherence function can be expanded as a power series in the fugacity using the single-particle propagator \cite{naraschewski_spatial_1999}. Starting from Eq.~\ref{eq: coherence function basic definition}, this procedure results in 
\begin{equation} \label{eq: coherence function}
    G^{(1)} \left( \bm{r}, \bm{r}' \right) = \frac{1}{ \lambda_T^2 } \sum_{j = 1}^\infty \frac{ z^j }{j} e^{ - \pi \left( \bm{r} - \bm{r}' \right)^2 / j \lambda_T^2 } 
\end{equation}
The first-order coherence function can be normalized by the (homogeneous) density $n = G^{(1)} \left( \bm{r}, \bm{r} \right) = - \log \left( 1 - z \right) / \lambda_T^2$. Taking the classical Maxwell-Boltzmann limit $z \to 0$ yields a Gaussian first-order coherence function with classical coherence length $\xi_0 = \lambda_T / \sqrt{\pi}$. In the following, we use ``Bose gas correlations'' to refer to a system with a finite fugacity and first-order coherence function given by Eq.~\ref{eq: coherence function}, as shown in Fig.~\ref{fig: correlations}a. We will primarily be concerned with the $z \to 1$ limit where the correlations deviate significantly from a Gaussian (see SM Sec.~\ref{sec: background on coherence}). 

The time-correlations can be studied in a similar manner to Eq.~\ref{eq: coherence function} after generalizing the definition of the first-order coherence function to finite time-separations. Following Ref.~\cite{kohnen_temporal_2015}, the (equal-position) first-order coherence function for a homogeneous system in two dimensions can be expressed in terms of the Lerch transcendent which goes asymptotically as
\begin{equation} \label{eq: coherence function time}
    G^{(1)} \left( t, t' \right) \sim \frac{1}{\lambda_T^2} \frac{z}{ 1 - z} \frac{ e^{ - i \tan^{-1} \left( \left( t - t' \right) / \theta \right)}}{ \sqrt{ 1 + \left( t - t' \right)^2 / \theta^2 }}
\end{equation}
in the limit $t - t' \gg \theta$ \cite{cai_note_2019}, defining the thermal time-scale $\theta = \hbar / k_B T$. The result is shown in Fig.~\ref{fig: correlations}b. The characteristic time-scale of the decay depends strongly on the fugacity $z$ and, consequently, the density $n$. For an EP system at cryogenic temperatures $T \sim 20$~K, the limit $t - t' \gg \theta$ is satisfied for time-separations much longer than approximately $350$~fs. 

The first-order coherence function is experimentally accessible in the intensity pattern of an interferometer which is determined by the visibility function $\mathcal{V}$. Under reasonable assumptions (see SM Sec.~\ref{sec: interference experiments}), the visibility function is related to the first-order coherence function as
\begin{equation} \label{eq: visibility}
    \mathcal{V}\left( \bm{r}, \tau \right) = \frac{ \text{Re} \left\{ G^{(1)} \left( \left( x, y, z \right), \left( -x, y, z \right), \tau \right) \right\} }{ G^{(1)} \left( \bm{r}, \bm{r} \right) }
\end{equation}
from which the traditional visibility $V$ measured from the contrast of interference fringes is derived $V = \left| \mathcal{V} \right|$. Optical fields with partial coherence $0 < V < 1$ diffract and interfere differently than their completely incoherent and completely coherent counterparts; however, the paraxial propagation of completely coherent optical fields in free space can be generalized to the quasi-monochromatic partially coherent case \cite{wolf_optical_1995, wolf_introduction_2007, gbur_structure_2010}, further simplifying in the far-field Fresnel and Fraunhofer approximations \cite{goodman_fourier_1968} (see SM Sec.~\ref{sec: interference experiments}). The diffraction of a source with Bose gas correlations is an interesting problem in its own right, which we explore in SM Sec.~\ref{sec: diffraction of partially coherent sources} (in particular SM Sec.~\ref{sec: diffraction with Bose gas correlations} and Fig.~\ref{fig:optics and diffraction}). 

Motivated by the definition of the condensate fraction \cite{penrose_bose-einstein_1956, sakmann_many-body_2011}, we define the coherent fraction of an optical field which is agnostic of the underlying phase (i.e. does not necessarily indicate BEC). In particular, we would like it to be measurable in interferometry experiments. Therefore, we define the coherent fraction $C$ of an optical field as
\begin{equation} \label{eq: coherent fraction}
     C = \frac{ \left| \int d^2r \; \text{Re} \left( G^{(1)} \left( \bm{r}, - \bm{r}, \tau \right) \right) \right| }{ \int d^2r \; G^{(1)} \left( \bm{r}, \bm{r}, 0 \right) }
\end{equation}
assuming rotational invariance. Importantly, the coherent fraction is conserved under paraxial propagation (see SM Sec.~\ref{sec: coherent fraction conservation}). While the (normalized) mutual intensity function provides a local measure of coherence, the coherent fraction provides a global measure of coherence for the entire optical field. 

\section{Behavior of the coherent fraction} \label{sec: results}

As a point of comparison, we begin by calculating the coherent fraction for free-propagating partially coherent optical field. We assume an ideal interferometer with zero time-delay $\tau = 0$, and an (equal-time) first-order coherence function which takes the generic form $G_0^{(1)} \left( \bm{s}_1, \bm{s}_2 \right) = \sqrt{ n \left( \bm{s}_1 \right) n \left( \bm{s}_2 \right) } g^{(1)} \left( \left| \bm{s}_1 - \bm{s}_2 \right| \right)$ where $n \left( \bm{s}_j \right)$ is the density at position $\bm{s}_j$. Specifically, we first consider a Gaussian-Schell model source \cite{collett_beams_1980, wolf_introduction_2007, gbur_structure_2010} so that the first-order coherence function in the plane $Z = 0$ with coordinates $\bm{s}_1$ and $\bm{s}_2$ is given by $G_0^{(1)} \left( \bm{s}_1, \bm{s}_2 \right) = \left| A \right|^2 e^{- \left( s_1^2 + s_2^2 \right) / \sigma^2 } e^{ - \left( \bm{s}_1 - \bm{s}_2 \right)^2 / \xi^2}$ where $A$ is the (complex) amplitude of the scalar field, $\sigma$ is the width of the Gaussian density profile, and $\xi$ is the coherence length. Simply integrating, the coherent fraction $C_{\text{GS}}$ measured over a disk of radius $a$ is then
\begin{equation} \label{eq: GS free coherent fraction}
    C_{\text{GS}} = \frac{1}{1 - e^{-2 a^2 / \sigma^2}} \frac{1}{1 + 2 \sigma^2 / \xi^2} \left( 1 - e^{- 2 a^2 / \sigma^2 - 4 a^2 / \xi^2} \right)
\end{equation}
The limit of uniform intensity $\sigma \to \infty$ with the aperture radius $a$ and coherence length $\xi$ fixed will be particularly relevant.

\begin{figure}[t]
    \centering
    \includegraphics[width=0.98\linewidth]{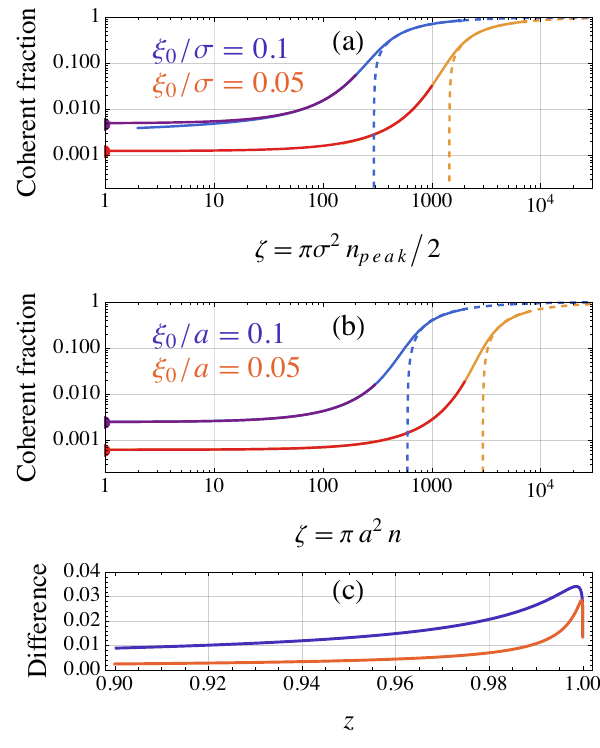}
    \caption{Ideal coherent fraction for a Gaussian beam with Bose gas correlations incident on a circular aperture as a function of (observed) particle number $\zeta$. (a) Coherent fraction for a Gaussian beam of width $\sigma$ in the limit of no aperture (see text) as a function of the particle number $\zeta = \pi \sigma^2 n_\text{peak} / 2$ measured in terms of the peak density $n_\text{peak}$. Two values of the classical coherence length relative to the beam width are shown, $\xi_0 / \sigma = 0.1$ (blue-purple) and $0.05$ (orange-red). For different regimes of particle number, different expression for the coherent fraction are used: Coherent fraction from series expansion in the fugacity Eq.~\ref{eq: BG free coherent fraction} (solid purple or red), coherent fraction from Euler-Maclaurin formula (solid blue or orange), and asymptotic behavior (dashed blue or orange) (see SM Sec.~\ref{sec: coherent fraction extended}). 
    (b) Coherent fraction for a beam incident on a circular aperture of radius $a$ in the limit of uniform intensity (see text) as a function of the particle number in the aperture $\zeta = \pi a^2 n$ measured in terms of the homogeneous density $n$. Expressions for the coherent fraction derived identically to those in (a) are used in (b) (see SM Sec.~\ref{sec: coherent fraction extended}).
    (c) Comparison of the coherent fractions from the limiting cases in (a) and (b) as a function of fugacity $z$. The difference is given by (a) - (b).
    }
    \label{fig:coherent fraction no diffraction}
\end{figure}

Second, we calculate the coherent fraction for a source with Bose gas correlations. In contrast to the Gaussian calculation, we fix $\xi = \xi_0$ at the classical value which no-longer plays the role of the coherence length and instead vary the fugacity $z$. The first-order coherence function in the plane $Z = 0$ with coordinates $\bm{s}_1$ and $\bm{s}_2$ is given by 
\begin{equation} \label{eq: full mif}
    G_0^{(1)} \left( \bm{s}_1, \bm{s}_2 \right) = \frac{ \left| A \right|^2 e^{- \left( s_1^2 + s_2^2 \right) / \sigma^2 } }{ - \log \left(1 - z \right) } \sum_{j = 1}^\infty \frac{ z^j }{j} e^{ - \left( \bm{s}_1 - \bm{s}_2 \right)^2 / j \xi_0^2} 
\end{equation}
where $A$ and $\sigma$ retain the same meaning. The coherent fraction $C_{\text{BG}}$ measured over a disk of radius $a$ is then
\begin{equation} \label{eq: BG free coherent fraction}
    C_{\text{BG}} = \frac{1}{- \log \left( 1 - z \right)} \sum_{j = 1}^\infty \frac{z^j}{j} C_{\text{GS}} \left( \sqrt{j} \xi_0 \right)
\end{equation}
where $C_{\text{GS}} \left( \sqrt{j} \xi_0 \right)$ denotes the Gaussian-Schell result Eq.~\ref{eq: GS free coherent fraction} evaluated at the adjusted coherence length $\sqrt{j} \xi_0$. (An alternative integral expression equivalent to Eq.~\ref{eq: BG free coherent fraction} is presented in SM Sec.~\ref{sec: coherent fraction momentum space}). The approximate behavior of Eq.~\ref{eq: BG free coherent fraction} can be obtained in various limits (see SM Sec.~\ref{sec: coherent fraction limits}), and the results are shown in Fig.~\ref{fig:coherent fraction no diffraction}. The coherent fraction after removing the aperture $a \to \infty$ with the width $\sigma$ and coherence length $\xi$ fixed is shown in Fig.~\ref{fig:coherent fraction no diffraction}a, while the coherent fraction in the limit of uniform intensity $\sigma \to \infty$ with the aperture radius $a$ and coherence length $\xi$ fixed is shown in Fig.~\ref{fig:coherent fraction no diffraction}b. Although qualitatively similar, the coherent fraction differs significantly for a fixed fugacity $z$, as shown in Fig.~\ref{fig:coherent fraction no diffraction}c. In particular, the coherent fraction in the limit of no aperture is always higher than the coherent fraction in the limit of uniform intensity since the aperture places a sharp cutoff on the points which can contribute to the coherence. 

The natural time-scale of the noninteracting, equilibrium Bose gas is the thermal time $\theta = \hbar / k_B T$, as introduced in Eq.~\ref{eq: coherence function time}. In EP systems, this should be compared to the inverse spectral width $1 / \Delta \nu$ of the monitored quasi-monochromatic optical field. At cryogenic temperatures $T \sim 20$~K, the thermal time is approximately $350$~fs, while the coherence time in the classical Maxwell-Boltzmann limit is $\pi \theta$ or approximately $1$~ps (see SM Sec.~\ref{sec: coherent fraction finite time-delay}). This is comparable to the inverse spectral width on the order of $1$~ps observed in experiment in the low-density limit \cite{alnatah_coherence_2024}. Notably, the optical signal from EP systems is known to exhibit spectral inhomogeneous broadening in the high-density regime \cite{timofeev_exciton_2012}. Due to the limited time-resolution, we expect there to be a small but finite time-delay in interferometry experiments.  

A finite time-delay generically reduces the strength of the equal-position correlations, as shown in Fig.~\ref{fig: correlations}b. However, the correlations of far-separated positions can actually become stronger at finite time-separations as the wavefunction evolves and propagates. This behavior combined with the real-part in the definition Eq.~\ref{eq: coherent fraction} has counterintuitive implications on global measurements of coherence, leading to non-monotonic and oscillatory behavior of the coherent fraction with increasing (classical) coherence length--equivalently decreasing temperature--and increasing time-separation (see SM Sec.~\ref{sec: coherent fraction finite time-delay}). The correlation area, on the other hand, is more well-behaved. We include a finite time-delay into the expression for the coherent fraction Eq.~\ref{eq: BG free coherent fraction} by replacing the Gaussian-Schell coherent fraction Eq.~\ref{eq: GS free coherent fraction} for zero time-delay $\tau = 0$ with the generalization for $\tau \neq 0$ (see SM Eq.~\ref{eq: BG free space-time coherent fraction}). The modified coherent fraction is shown in Fig.~\ref{fig: theory experiment comparison}a for a range of time-delays $\tau$ between interferometry arms, see Sec.~\ref{sec: discussion} for further discussion. 

\section{Discussion and comparison to experiment} \label{sec: discussion}

Our analysis considers an equilibrium gas of noninteracting EPs in a homogeneous potential landscape. In this section, we discuss the validity of these assumptions and compare to recent experimental results \cite{alnatah_coherence_2024}.

First, we neglect nonequilibrium effects, and our analysis presumes an equilibrated thermal state rather than a non-equilibrium steady-state. Measurements of the thermal distribution function in Ref.~\cite{alnatah_coherence_2024} show good agreement with the Bose-Einstein distribution suggesting equilibrium conditions. Although, the long-wavelength dynamics of EP systems in one and two dimensions are ultimately governed by the KPZ equation \cite{diessel_emergent_2022, fontaine_kardarparisizhang_2022} (complicated in two dimensions by BKT behavior at shorter length-scales  \cite{wachtel_electrodynamic_2016}), for finite-sized EP systems the phenomena of quasi-condensation applies: approximately uniform phase and constant long-range correlations across the sample.

\begin{figure}
    \centering
    \includegraphics[width=0.98\linewidth]{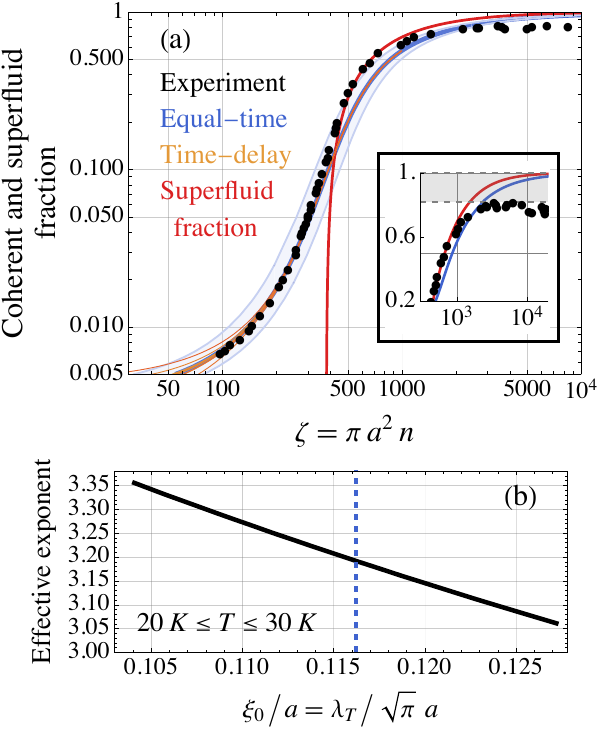}
    \caption{Comparison of the noninteracting, equilibrium theory presented in the main text to recent experimental results from Ref.~\cite{alnatah_coherence_2024}. 
    (a) The experimental coherent fraction measurements (black circles) are plotted in terms of the EP number in the aperture $\zeta = \pi a^2 n$. The theory prediction for the coherent fraction measured by an ideal interferometer with zero time-delay $\tau = 0$ (solid dark blue) as given by Eq.~\ref{eq: BG free coherent fraction} and its extensions (see SM Sec.~\ref{sec: coherent fraction extended}) plotted for a range of temperatures $21 \leq T \leq 27$~K with $T = 24$~K the center. 
    The coherent fractions measured by an interferometer with finite time-delays $2 \leq \tau \leq 6$~ps (multiple solid orange) is plotted at temperature $T = 24$~K (see SM Sec.~\ref{sec: background on coherence} and \ref{sec: coherent fraction extended}). 
    The superfluid fraction (solid red) defined in Eq.~\ref{eq: superfluid fraction} using $g = 5$~$\mu$eV-$\mu$m$^2$ matches the near-transition data.
    (Inset) The experimental coherent fraction measurements saturate below $1$ at high densities.
    (b) Effective power-law exponent characterizing the growth of the coherent fraction as measured by an ideal interferometer with zero time-delay as a function of the classical coherence length relative to the aperture radius $\xi_0 / a$ (see text). The values of $\xi_0 / a$ shown correspond to temperatures $20$~K $\leq T \leq$~$30$~K for the conditions assumed in (a) and predict an effective power-law exponent of approximately $3.19$ (dashed dark blue) at $T = 24$~K. All plots use the EP mass $m = 1.515 \times 10^{-4} m_e$, where $m_e = 9.109 \times 10^{-31}$~kg is the vacuum electron mass, and the aperture radius $a = 6$~$\mu$m. 
    }
    \label{fig: theory experiment comparison}
\end{figure}

Second, we neglect interaction effects. Two-body interactions of EP are generically considered to be weak, yet they can still play a crucial role in the system's correlations and phase, for example facilitating the BKT phase transition. 
While the relation between $g$, the interaction strength, and $b$, the interaction length-scale, is obscured in two-dimensions
\cite{schick_two-dimensional_1971, lieb_rigorous_2001, posazhennikova_colloquium_2006}, to estimate the length-scale of perturbations induced by the interaction as observed in the first-order coherence function after the aperture, we compute the interaction energy $g n$, relate it to the momentum $k_g = \sqrt{ 2 m g n / \hbar^2}$, then extract the length-scale $b = Z k_g / \bar{k}$ motivated by the far-field position variable arising in the paraxial propagation. From the values of $g$, $n$, and $\bar{k}$ provided in Ref.~\cite{alnatah_coherence_2024}, we find the ratio $k_g / \bar{k} \approx 0.04 \sim 10^{-2}$. 
However, we note that the measurement of $g$ in EP systems is difficult, and $g$ is often treated as a free parameter. For one, excitonic disorder, modeled as a distribution of excitonic energies and photon-coupling strengths \cite{marchetti_absorption_2007}, can be effectively described by a significantly enhanced two-body interaction strength in the low-density regime \cite{marchetti_phase_2008}. 
But overall, since this ratio is small, we expect that the interactions do not dramatically alter the first-order coherence function, and that it is sufficient to approximate the EP gas as noninteracting away from the critical density. 

With these notes in mind, we compare the predictions of the finite-sized, noninteracting, equilibrium theory to the recent experimental results from Ref.~\cite{alnatah_coherence_2024} in Fig.~\ref{fig: theory experiment comparison}. The model for an ideal interferometer with zero time-delay $\tau = 0$ has no free parameters; the only inputs are the EP mass $m = 1.515 \times 10^{-4} m_e$, where $m_e = 9.109 \times 10^{-31}$~kg is the vacuum electron mass, the temperature $T \simeq 24$~K, and the aperture radius $a$, entering through the dimensionless classical coherence length $\xi_0 = \lambda_T / \sqrt{\pi} a$. The data shown in Fig.~\ref{fig: theory experiment comparison}a is for an aperture radius $a = 6$~$\mu$m. Experimental measurements of the temperature show significant fluctuations, and so we include theory curves for a range of temperatures $21 \leq T \leq 27$~K in Fig.~\ref{fig: theory experiment comparison}a which bound the experimental data. This point notwithstanding, the curve corresponding to $T = 24$ shows good agreement with the experimental data in the low-density regime. 

Next, we extract effective power-law exponents characterizing the growth of the coherent fraction as a function of the density (via the EP number in the aperture) near the transition by fitting the slope of the coherent fraction relative to the zero-density limit as a function of EP number in the aperture on a log-log scale at the inflection point, occurring in the region of steepest growth. We stress, however, that the buildup of coherence is not formally described by a power-law. As shown in Fig.~\ref{fig: theory experiment comparison}b, the theory predicts faster growth as $\xi_0 / a$ is decreased and slower growth as $\xi_0 / a$ is increased, a result of tuning the relative size of the EP sample. At the temperature $T = 24$~K with the parameters given in the previous paragraph, the predicted effective power-law exponent is approximately $3.19$, consistent with the effective power-law exponent obtained in Ref.~\cite{alnatah_coherence_2024}. The other values of $\xi_0 / a$ correspond to temperatures in the range $20$~K $\leq T \leq$~$30$~K, consistent with those observed in Ref.~\cite{alnatah_coherence_2024}.

As a comparison to the zero time-delay predictions, we also show predictions for a small range of time-delays $\tau = 2$~ps to $6$~ps at temperature $T = 24$~K in Fig.~\ref{fig: theory experiment comparison}a. The uncertainty in the experimental time-delay measurement $\delta \tau$ can be estimated from the uncertainty of the path-length difference between the interferometry arms, approximately $2$~ps in Ref.~\cite{alnatah_coherence_2024}. Below $\tau = 2$~ps, there is no effect of time-delay at the scale shown due to the exponential suppression of corrections when $\tau \ll \theta a / \xi_0$ (see SM Sec.~\ref{sec: coherent fraction finite time-delay}). 
As the time-delay increases, the expression is notably non-monotonic. However, at high densities, the result for finite time-delay becomes coincident with the $\tau = 0$ result, as shown in Fig.~\ref{fig: theory experiment comparison}a.

Near the transition, we compare the coherent fraction to the superfluid fraction, which exists only for interacting gases. In terms of the parameters used in this paper, we calculate the superfluid fraction $s$ (ignoring the finite size) as \cite{hohenberg_microscopic_1965, posazhennikova_colloquium_2006}
\begin{equation} \label{eq: superfluid fraction}
    s = 1 - \frac{1}{\zeta \xi_0^2 / a^2} \int_0^\infty d x \left( 1 - \frac{1}{ \sqrt{ 1 + \frac{x^2}{\gamma^2 \zeta^2 \xi_0^4  / a^4}}} \right) \frac{ x e^x }{ \left( e^x - 1 \right)^2 }
\end{equation}
where $\gamma = m g / 2 \pi \hbar^2$ is a dimensionless parameter characterizing the interaction strength. 
The superfluid fraction is compared to the coherent fraction in Fig.~\ref{fig: theory experiment comparison}a showing improved agreement near the transition using the interaction strength $g = 5$~$\mu$eV-$\mu$m$^2$, consistent with the value found in Ref.~\cite{alnatah_coherence_2024} within the experimental uncertainty. 

This comparison has two important implications. First, the superfluid fraction can be roughly identified with the coherent fraction when the Bose gas is in the superfluid state, but not when it is in the normal state. Although the coherent fraction varies continuously through the transition, the normal state exhibits only short-range correlations while the superfluid state exhibits long-range correlations. (This crossover of the long-range behavior is seen clearly in simulations using the Gross-Pitaevskii equation in Ref.~\cite{alnatah_coherence_2024}). 
Second, the fit of the superfluid fraction to the near-transition coherence data functions as a measurement of the effective interaction strength $g$. We find that the fit constrains the interaction strength $g$ to a range $1.5-10$ $\mu$eV-$\mu$m$^2$, noting that the fitted value varies with the temperature. This range is well above the theoretical prediction $g \sim 0.25~\mu$eV-$\mu$m$^2$ for polariton-polariton interactions in the case of no free excitons, verified experimentally, but well below the value of $g \sim 40~\mu$eV-$\mu$m$^2$ estimated for polariton-exciton interactions \cite{snoke_reanalysis_2023}. This result is somewhat surprising since Ref.~\cite{alnatah_coherence_2024} studied polaritons coexisting with a spatially homogeneous exciton population 200-1000 times more dense. The observed coherence is, therefore, not a direct result of polariton-exciton interaction, but instead a result of effective polariton-polariton interactions influenced by the presence of the excitons.

The noninteracting theory and interacting superfluid theory both predict that the coherent and superfluid fractions, respectively, approach exactly 1 in the high-density limit. The experimental measurements, however, never measure complete coherence. It is likely that the interferometer in the experiment has systematic uncertainties that prevent measurement of perfect contrast of the fringes. Alternatively, a full many-body theory might give depletion of the coherent fraction at high density.

\section{Conclusion}

The results presented in this paper are broadly applicable to interference experiments with partially coherent light; in particular, they may be helpful in measurements of coherence in finite-sized EP systems characterized by Gaussian or more general Bose gas correlations. For EP systems, our results provide a baseline, without fit parameters, to which interacting models and experimental data can be compared. 

In particular, we find good agreement to the data of Ref.~\cite{alnatah_coherence_2024} over nearly three orders of magnitude in the coherent fraction and nearly a factor of 50 in the EP density at approximately constant temperature. The noninteracting theory works well in the normal state, while the standard theory of a weakly-interacting superfluid works well near the critical density. It is somewhat surprising that the noninteracting theory works so well, since it is well known that the EP at all densities coexist with a homogeneous gas of excitons which have much higher density. This indicates that, although the EP gas is affected by the presence of the excitons in various ways, namely by the shift of their ground-state energy and collisional line broadening \cite{snoke_reanalysis_2023}, it can still be treated as a weakly-interacting, equilibrium Bose gas to a high approximation.

\begin{acknowledgments}
This research is funded in part by the Gordon and Betty Moore Foundation through grant GBMF12763 to Peter Littlewood, who thanks Shuolong Yang for discussions. The experimental work at Pittsburgh and sample fabrication at Princeton were supported by the National Science Foundation through grant DMR-2306977. 
\end{acknowledgments}


\bibliography{bibliography}

\widetext
\clearpage
\begin{center}
\textbf{\large Supplemental Material: Optical probes of coherence in two dimensional Bose gases of polaritons}
\end{center}
\setcounter{section}{0}
\setcounter{equation}{0}
\setcounter{figure}{0}
\setcounter{table}{0}
\setcounter{page}{1}
\makeatletter
\renewcommand{\thesection}{S\arabic{section}}
\renewcommand{\theequation}{S\arabic{equation}}
\renewcommand{\thefigure}{S\arabic{figure}}
\renewcommand{\bibnumfmt}[1]{[S#1]}
\renewcommand{\citenumfont}[1]{S#1}

In this supplement, we provide theoretical background for the calculations in the main text, expand upon main results, as well as introduce some supplemental analysis. The first three sections are focused on background. In Sec.~\ref{sec: background on coherence}, we introduce the noninteracting EP system and study it via the first-order coherence function. Moreover, we explain the limiting behavior of the first-order coherence function and discuss its generalization to finite time-separations. This introduction allows us to discuss the measurement of coherence in EP systems via optical interferometry in Sec.~\ref{sec: interference experiments}. We also include a discussion of the diffraction of partially coherent sources which will be important in Sec.~\ref{sec: diffraction of partially coherent sources}. Next, in Sec.~\ref{sec: coherent fraction definition}, we revisit the definition of the coherent fraction and show that it is conserved under paraxial propagation in Sec.~\ref{sec: coherent fraction conservation}, a result cited in the main text. 

The final three sections expand upon main results and introduce some supplemental analysis. In Sec.~\ref{sec: coherent fraction extended} we provide an extended analysis of the coherent fraction, in particular the expressions Eq.~\ref{eq: GS free coherent fraction} and Eq.~\ref{eq: BG free coherent fraction} in the main text. Then, in Sec.~\ref{sec: diffraction of partially coherent sources}, we explore the diffraction of partially coherent sources with Gaussian and Bose gas correlations, summarized in Fig.~\ref{fig:optics and diffraction}, and provide explicit formulae for the propagation integrals in Sec.~\ref{sec: propagation integrals}. Lastly, in Sec.~\ref{sec: additional data and GPE}, we analyze additional data in Fig.~\ref{fig: extended data} supplementing Fig.~\ref{fig: theory experiment comparison} of the main text and make the connection to GPE simulation clear, as discussed in Ref.~\cite{alnatah_coherence_2024}.

\section{Theory of coherence in Bose gases} \label{sec: background on coherence}

In this first section, we review the microscopic description of EP systems as noninteracting Bose gases, and discuss application to the first-order coherence function as introduced in Sec.~\ref{sec: theory of Bose gases} of the main text. The Hamiltonian for a gas of (lower) EPs in second-quantized form is
\begin{equation}
    H = \sum_{\bm{k}_\parallel} \epsilon \left( \bm{k}_\parallel \right) P_{\bm{k}_\parallel}^\dagger P_{\bm{k}_\parallel}
\end{equation}
where $\bm{k}_\parallel$ is the in-plane component of the EP momentum perpendicular to the cavity axis, $\epsilon \left( \bm{k}_\parallel \right)$ is the in-plane EP dispersion, and $P_{\bm{k}_\parallel} = A_{\bm{k}_\parallel} a_{\bm{k}_\parallel} + B_{\bm{k}_\parallel} b_{\bm{k}_\parallel}$ is the EP annihilation operator constructed from the photon annihilation operator $a_{\bm{k}_\parallel}$, exciton annihilation operator $b_{\bm{k}_\parallel}$, and Hopfield coefficients $A_{\bm{k}_\parallel}$ and $B_{\bm{k}_\parallel}$ which depend on the cavity detuning and the exciton-photon coupling. Since cavity photons have finite lifetimes, the EP wavefunction is accessible as a projection onto the escaped photons $a_{\bm{k}_\parallel}$.

The (equal-time) first-order coherence function of the EP system described above is captured by Eq.~\ref{eq: coherence function basic definition} and \ref{eq: coherence function}. The corresponding (equal-time) normalized first-order coherence function $g^{(1)} \left( r \right)$ is plotted in Fig.~\ref{fig: correlations}a for various fugacities. For small fugacities $z \ll 1$, it is useful to consider an effective Gaussian first-order coherence function characterized by an effective coherence length $\xi_{\text{eff}}$ \cite{corman_two-dimensional_2016}. An example Gaussian fit to the exact first-order coherence function at $z = 0.2$ is shown in Fig.~\ref{fig: correlations}a. 

The deviation of the exact first-order coherence function from the Gaussian form becomes more prominent as the fugacity increases. Moreover, we are primarily interested in the limit $z \to 1$. The long-range behavior of the first-order coherence function is determined by the low-momentum behavior of the Bose-Einstein distribution \cite{hadzibabic_two-dimensional_2011, saint-jalm_exploring_2019}. Expansion of the exponential in powers of $k_\parallel$ yields
\begin{equation}
    \tilde{n} \left( \bm{k}_\parallel \right) = \frac{z}{e^{\beta k_\parallel^2 / 2m} - z} = \frac{z}{1 + \beta k_\parallel^2 / 2 m + \mathcal{O} \left( k_\parallel^4\right) - z}
\end{equation}
After inverse Fourier transform of the rational function, we obtain the asymptotic form of the first-order coherence function as a function of the position separation $r = \left| \bm{r} - \bm{r}' \right|$
\begin{equation} \label{eq: coherence function asymptotic}
    G^{(1)} \left( r \right) \sim \frac{2 z}{\lambda_T^2} K_0 \left( \frac{2 \sqrt{ \pi} \sqrt{ 1 - z} \: r }{ \lambda_T} \right)
\end{equation}
where $K_0$ is the zeroth-order modified Bessel function of the second kind. Since this result works best in the limit $z \to 1$, it is also common to expand the fugacity $z = 1 + \beta \mu + \mathcal{O}\left(\mu^2\right)$ in terms of the chemical potential in the limit $\mu \to 0^-$. 

We can estimate the fugacity $z$ at which Bose gas correlations dominate. The length-scale $\ell$ associated with the asymptotic first-order coherence function is 
\begin{equation}
    \ell = \frac{\lambda_T}{ 2 \sqrt{\pi} \sqrt{ 1 - z } } \sim \frac{1}{ \sqrt{ - 2 m \mu }}
\end{equation}
and, therefore, the length $\ell$ surpasses the classical coherence length $\xi_0 = \lambda_T / \sqrt{\pi}$ at $z = 3/4$.

The discussion above can be generalized to finite time-separations, as mentioned in the main text. Following Ref.~\cite{kohnen_temporal_2015}, the full first-order coherence function as a function of the position separation $r = \left| \bm{r} - \bm{r}' \right|$ and the time-separation $\tau = t - t'$ is
\begin{equation} \label{eq: space-time first-order coherence function}
    G \left( r, \theta \right) = \frac{1}{\pi \xi_0^2} \sum_{j = 1}^\infty \frac{z^j}{j + i \tau / \theta} e^{ - r^2 / \xi_0^2 \left( j + i \tau / \theta \right) }
\end{equation}
recalling the thermal time-scale $\theta = \hbar / k_B T$. The (equal-position) first-order coherence function provided in Eq.~\ref{eq: coherence function time} corresponds to the special case $G \left( 0, \tau \right)$ for $\tau \gg \theta$. The corresponding (equal-position) normalized first-order coherence function $g^{(1)} \left( \tau \right)$ is plotted in Fig.~\ref{fig: correlations}b for various fugacities. The asymptotic form Eq.~\ref{eq: coherence function time} converges to the exact result for time-separations $\tau$ satisfying $\tau \gg \theta$, also shown in Fig.~\ref{fig: correlations}b.

\section{Interferometry and diffraction with partial coherence} \label{sec: interference experiments}

The first-order coherence function is experimentally accessible in the intensity pattern of an interferometer. EP systems well below the BEC transition are approximately unpolarized, and therefore the illuminating source field can be treated as a scalar $u_S$. We further assume that $u_S$ is stationary and ergodic. For a balanced Michelson interferometer which inverts in the $x$-direction and an intensity profile with (at least) $x$-symmetry, the output intensity is
\begin{equation}
    \left\langle I \left( \bm{r} \right) \right\rangle = \frac{1}{2} \left\langle I_S \left( \bm{r} \right) \right\rangle \left( 1 + \mathcal{V}\left(\bm{r}, \tau \right) \right)
\end{equation}
where $I_S = \epsilon_0 c \left| u_S \right|^2 / 2$ is the source intensity, $\epsilon_0$ is the permittivity of free-space, $c$ is the speed of light in vacuum, and $\tau = t - t'$ is the time-delay between interferometry arms, set by the path-length difference. The brackets $\left\langle \cdot \right\rangle$ denote the time (ensemble) average in the classical (quantum) regime. The dependence on the time-delay $\tau$ may be very strong for light with a short coherence time. The visibility function $\mathcal{V}$ is related to the first-order coherence function as
\begin{equation}
    \mathcal{V}\left( \bm{r}, \tau \right) = \frac{\epsilon_0 c}{ 2 \left\langle I_S \left( \bm{r} \right) \right\rangle} \text{Re} \left\{ G^{(1)} \left( \left( x, y, z \right), \left( -x, y, z \right), \tau \right) \right\} 
\end{equation}
equivalent to Eq.~\ref{eq: visibility} in the main text, from which the traditional visibility $V$ measured from the contrast of interference fringes is derived $V = \left| \mathcal{V} \right|$. The visibility satisfies $0 \leq V \leq 1$. In the limit of complete coherence, $V = 1$, the contrast of the interference fringes is maximized, ranging from the maximum intensity to zero intensity. In the limit of complete incoherence (although physically the coherence length is lower-bounded by the light wavelength \cite{wolf_introduction_2007}), $V \approx 0$ and the contrast of the interference fringes is minimized.  

A quasi-monochromatic beam has a small but finite spectral width $\Delta \nu$ and can exhibit interesting spectral anomalies near singular points, points of zero intensity \cite{gbur_anomalous_2001, gbur_singular_2002, ponomarenko_spectral_2002}. The finite spectral width is a result of a finite coherence time which suggests that the first-order coherence function may depend strongly on the time-delay $\tau$. But so long as the time-delay $\tau$ is much smaller than the inverse of the spectral width $1 / \Delta \nu$, then its effect can be approximated as a phase factor in the first-order coherence function \cite{wolf_optical_1995}. In the main text, we primarily focus on the spatial behavior of the first-order coherence function and often assume the time-delay to be optimized $\tau = 0$, unless otherwise stated.

In this section and some others which follow, we use the conventional optics nomenclature for ease of comparison. (1) The mutual coherence function $\Gamma$ is the first-order coherence function $G^{(1)}$ of the optical field. (2) The mutual intensity function is the mutual coherence function for zero time-delay $\tau = 0$ (i.e. the equal-time first-order coherence function). (3) The intensity is defined via the mutual intensity function $I \left( \bm{r} \right) = \epsilon_0 c \: \Gamma \left(  \bm{r},  \bm{r} \right) / 2$ (similarly to the particle density in a massive gas). With this definition, we can omit the brackets $\left\langle \cdot \right\rangle$ since the mutual intensity function is already averaged. 

The paraxial propagation of completely coherent optical fields in free space can be generalized to the quasi-monochromatic partially coherent case \cite{wolf_optical_1995, wolf_introduction_2007, gbur_structure_2010}. Given a mutual coherence function $\Gamma_0$ in the plane $Z = 0$ with coordinates $\bm{s}_1$ and $\bm{s}_2$, the propagated mutual intensity function in a plane of constant $Z$ with coordinates $\bm{r}_1$ and $\bm{r}_2$ is calculated via an integral transform. In particular, the far-field Fresnel and Fraunhofer approximations \cite{goodman_fourier_1968} yield \cite{wolf_optical_1995}
\begin{align}
    \Gamma_{\text{Fresnel}} \left( \bm{r}_1, \bm{r}_2, Z, \tau \right) & = \frac{1}{\bar{\lambda}^2 Z^2} e^{i \frac{\bar{k}}{2 Z} \left( r_1^2 - r_2^2 \right) } \; \mathfrak{F} \left\{ \Gamma_0 \left( \bm{s}_1, \bm{s}_2, \tau \right) P \left( \bm{s}_1 \right) P^* \left( \bm{s}_2 \right) e^{ i \frac{\bar{k}}{2 Z} \left( s_1^2 - s_2^2 \right) } \right\} \left( \bm{f}_1 = \frac{\bar{k} \bm{r}_1}{Z}, \bm{f}_2 = \frac{\bar{k} \bm{r}_2}{Z} \right) \label{eq: fresnel} \\
    \Gamma_{\text{Fraunhofer}} \left( \bm{r}_1, \bm{r}_2, Z, \tau \right) & = \frac{1}{\bar{\lambda}^2 Z^2} e^{i \frac{\bar{k}}{2 Z} \left( r_1^2 - r_2^2 \right) } \; \mathfrak{F} \left\{ \Gamma_0 \left( \bm{s}_1, \bm{s}_2, \tau \right) P \left( \bm{s}_1 \right) P^* \left( \bm{s}_2 \right) \right\} \left( \bm{f}_1 = \frac{\bar{k} \bm{r}_1}{Z}, \bm{f}_2 = \frac{\bar{k} \bm{r}_2}{Z} \right) \label{eq: fraunhofer}
\end{align}
where $\mathfrak{F} \left\{ \cdot \right\}$ is the four-dimensional Fourier transform with respect to the alternating kernel $e^{- i \left( \bm{f}_1 \cdot \bm{s}_1 - \bm{f}_2 \cdot \bm{s}_2 \right) }$ to the far-field ``momentum'' variables $\bm{f}_1 = \bar{k} \bm{r}_1 / Z$ and $\bm{f}_2 = \bar{k} \bm{r}_2 / Z$, $\bar{\lambda}$ is the mean wavelength of the quasi-monochromatic light, $\bar{k} = 2 \pi / \bar{\lambda}$ is the mean wavenumber, and $P$ is the (possibly complex) transmission of the aperture at $Z = 0$. While subsequent sections use the Fraunhofer approximation Eq.~\ref{eq: fraunhofer}, the Fresnel approximation Eq.~\ref{eq: fresnel} is necessary, e.g., for larger aperture radii. 

The difficulty of the propagation is rooted in the functional form of the correlations at $Z = 0$ and the effect of the aperture as modeled by the aperture transmission function $P$. But various simplifications of this calculation exist. For one, the appearance of a Fourier transform suggests application of the convolution theorem. Additionally, for isotropic correlations which only depends on the magnitude of the separation $\left| \bm{s}_1 - \bm{s}_2 \right|$, the intensity can be calculated with Schell's theorem \cite{schell_multiple_1961, nugent_generalization_1990}. Numerically, the existence of fast Fourier transform algorithms and parallel computation may aid calculations, as discussed in Ref.~\cite{shen_fast-fourier-transform_2006} for the completely coherent case. However, it should be noted that the completely coherent case only requires a two-dimensional integral rather than a four-dimensional integral.

\section{Coherent fraction definition} \label{sec: coherent fraction definition}

As a motivation for the definition of the coherent fraction discussed in the main text, we recall the definition of the condensate fraction $n_0$. Loosely, the condensate fraction is given by the long-range limiting value of the normalized first-order coherence function \cite{penrose_bose-einstein_1956}. More formally, the general expression in $d$-dimensions is
\begin{equation}
    n_0 = \frac{ \int d^d r_1 d^d r_2 \; \phi_0^* \left( \bm{r}_1 \right) G^{(1)} \left( \bm{r}_1, \bm{r}_2 \right) \phi_0 \left( \bm{r}_2 \right) }{ \int d^d r \; G^{(1)} \left( \bm{r}, \bm{r} \right) }
\end{equation}
where $\phi_0$ is the eigenfunction of $G^{(1)}$ with the largest occupation eigenvalue \cite{sakmann_many-body_2011}. In the noninteracting case, $\phi_0$ is simply the single-particle eigenfunction of the Hamiltonian with the largest occupation, the ground-state. Additionally, homogeneity implies $\phi_0$ is constant and $G^{(1)}$ only depends on the magnitude of the separation vector $r = \left| \bm{r}_1 - \bm{r}_2 \right|$, as seen before.

The coherent fraction $C$ of an optical field is, therefore, defined similarly
\begin{equation}
     C = \frac{ \left| \int d^2r \; \text{Re} \left( G^{(1)} \left( \bm{r}, - \bm{r}, \tau \right) \right) \right| }{ \int d^2r \; G^{(1)} \left( \bm{r}, \bm{r}, 0 \right) } 
\end{equation}
assuming rotational invariance. The integrand of the numerator is simply proportional to $I \left( \bm{r} \right) \mathcal{V} \left( \bm{r}, \tau \right)$. If the first-order coherence function is not rotational invariant, then the visibility needs to be measured for all orientations of the interferometer, in principle. The coherent fraction is bounded $0 \leq C \leq 1$ and conserved under propagation by Eq.~\ref{eq: fresnel} and Eq.~\ref{eq: fraunhofer}, as discussed in the subsequent section.

The principal difference between the definition of the coherent fraction $C$ and the condensate fraction $n_0$ is the real part in the numerator. This adjustment ensures that the coherent fraction is directly measurable in interferometry experiments through the contrast of interference fringes, even though nonzero time-delays generically produce complex-valued correlations. Note, the imaginary part can, in principle, be extracted via Hilbert transform relations \cite{wolf_optical_1995}. Practically, the integration in Eq.~\ref{eq: coherent fraction} can only be performed over a finite range which produces small corrections to the coherent fraction so long as the mutual coherence function is sufficiently localized. However, the localization condition can be broken in the case of an optical field diffracting through a small aperture.

\subsection{Conservation of the coherent fraction under propagation} \label{sec: coherent fraction conservation}

The definition of the coherent fraction presented in Eq.~\ref{eq: coherent fraction} is conserved under Fresnel and Fraunhofer propagation, Eq.~\ref{eq: fresnel} and Eq.~\ref{eq: fraunhofer} respectively. This result is derived as follows. Consider a first-order coherence function (mutual intensity function) of the generic form
\begin{equation}
    G^{(1)}_0 \left( \bm{s}_1, \bm{s}_2 \right) = \sqrt{ n \left( \bm{s}_1 \right) n \left( \bm{s}_2 \right) } g^{(1)} \left( \left| \bm{s}_1 - \bm{s}_2 \right| \right)
\end{equation}
where $n$ is the density and $g^{(1)}$ is the normalized first-order coherence function which is assumed to have rotational invariance.The coherent fraction is then
\begin{equation} \label{eq: coherent fraction initial}
    C = \frac{ \left| \int d^2s \; \sqrt{ n \left( \bm{s} \right) n \left( - \bm{s} \right) } \text{Re} \left( g^{(1)} \left( 2 s \right) \right) \right| }{ \int d^2s \; n \left( \bm{s} \right) }
\end{equation}

Now we introduce an aperture in the plane $Z = 0$ described by a transmission function $P$ and propagate the first-order coherence function with Eq.~\ref{eq: fresnel} to a plane of constant $Z$ with coordinates $\bm{r}_1$ and $\bm{r}_2$, with Eq.~\ref{eq: fraunhofer} being a limiting case. Applying Eq.~\ref{eq: coherent fraction} we find
\begin{equation} \label{eq: coherent fraction propagated}
    C = \frac{ \left|  \text{Re} \int d^2 s_1 d^2 s_2 \; \sqrt{ n \left( \bm{s}_1 \right) n \left( \bm{s}_2 \right) }  g^{(1)} \left(  \left| \bm{s}_1 - \bm{s}_2 \right| \right) P \left( \bm{s}_1 \right) P^* \left( \bm{s}_1 \right) e^{i \frac{\bar{k}}{2 Z} \left( s_1^2 - s_2^2 \right) } \int_W d^2r \; e^{- i \frac{\bar{k}}{Z} \bm{r} \cdot \left( \bm{s}_1 + \bm{s}_2 \right) } \right| }{ \int d^2 s_1 d^2 s_2 \; \sqrt{ n \left( \bm{s}_1 \right) n \left( \bm{s}_2 \right) } g^{(1)} \left(  \left| \bm{s}_1 - \bm{s}_2 \right| \right) P \left( \bm{s}_1 \right) P^* \left( \bm{s}_1 \right) e^{i \frac{\bar{k}}{2 Z} \left( s_1^2 - s_2^2 \right) } \int_W d^2r \; e^{- i \frac{\bar{k}}{Z} \bm{r} \cdot \left( \bm{s}_1 - \bm{s}_2 \right) }}
\end{equation}
where $W$ is the window in the plane of constant $Z$ over which the coherent fraction is measured. For $W = \mathbb{R}^2$ we obtain Eq.~\ref{eq: coherent fraction initial} (assuming integration only over the aperture) since the integrals over $\bm{r}$ yield Dirac $\delta$-functions. Therefore, the coherent fraction is conserved for all $Z$. 

We note that an alternate definition of the coherent fraction with $\left| \text{Re} \left( G^{(1)} \left( \bm{r}, - \bm{r}, \tau \right) \right) \right|$ in the numerator integrand of Eq.~\ref{eq: coherent fraction} could be used. This definition gives the ratio of the power of the optical field producing interference fringes to the total power. However, this definition is not, in general, conserved under propagation, although it is lower-bounded by the coherent fraction in the plane $Z = 0$.

\section{Extended coherent fraction calculations} \label{sec: coherent fraction extended}

In this section of the supplement, we expand upon the coherent fraction calculations presented in the main text. In particular, we provide an alternative equivalent expression to Eq.~\ref{eq: BG free coherent fraction} in Sec.~\ref{sec: coherent fraction momentum space} which is useful in some contexts. Then, in Sec.~\ref{sec: coherent fraction finite time-delay}, we provide details on the coherent fraction calculation at finite time-delay which appear in Fig.~\ref{fig: theory experiment comparison}. Finally, we discuss the behavior of the coherent fraction in various limits, with Sec.~\ref{sec: coherent fraction limits} particularly relevant to the results in the main text. 

\subsection{Alternative integral expression for the coherent fraction at zero time-delay} \label{sec: coherent fraction momentum space}

The first-order coherence function in position-space is presented in Eq.~\ref{eq: coherence function} which eventually leads to the coherent fraction results in Sec.~\ref{sec: results}. Here, we show that, starting from the first-order coherence function in momentum-space (i.e. with Eq.~\ref{eq: coherence function basic definition} and the Bose-Einstein momentum distribution), a relatively simple integral expression for the coherent fraction can be obtained. The calculation differs from that presented in SM Sec.~\ref{sec: propagation integrals} since we bypass the intermediate calculation of the propagated mutual intensity function. We stress that the expression presented here is equivalent to Eq.~\ref{eq: BG free coherent fraction} but with advantages and disadvantages: While Eq.~\ref{eq: BG free coherent fraction} allows for careful treatment of limits, as discussed in detail in SM Sec.~\ref{sec: coherent fraction limits}, the approach presented here directly connects to simulation with the Gross-Pitaevskii equation as described in Ref.~\cite{alnatah_coherence_2024}. 

Recall the (equal-time) first-order coherence function
\begin{equation}
    G^{(1)} \left( \bm{r}, \bm{r}' \right) = \int \frac{d^2 k_\parallel}{\left(2 \pi \right)^2} \; e^{i \bm{k}_\parallel \cdot \left( \bm{r} - \bm{r}' \right)} \tilde{n} \left( \bm{k}_\parallel \right)
\end{equation}
As in the main text, we integrate over a disk of radius $a$ so that by Eq.~\ref{eq: coherent fraction} the coherent fraction is
\begin{equation}
    C = \frac{1}{- \log \left( 1 - z \right)} \frac{\xi_0^2}{a^2} \int_D d^2 s \int \frac{d^2 k_\parallel}{\left(2 \pi \right)^2} \; e^{i 2 \bm{k}_\parallel \cdot \bm{s}} \tilde{n} \left( \bm{k}_\parallel \right)
\end{equation}
where $D$ denotes the disk region. We can perform the integral over the real-space coordinate $\bm{s}$ using Bessel functions as well as the integral over the angular position of the in-plane momentum $\bm{k}_\parallel$ to obtain
\begin{equation}
    C = \frac{1}{- \log \left( 1 - z \right)} \frac{\xi_0^2}{2 a^2} \int_0^\infty d \tilde{k} \; \tilde{n} \left( \tilde{k} \right) J_1 \left( 2 \tilde{k} \right)
\end{equation}
where $\tilde{k} = k_\parallel a$ is the dimensionless in-plane momentum. This expression produces equivalent predictions to the expressions in the main text, particularly Eq.~\ref{eq: BG free coherent fraction} and its extensions, as presented in Fig.~\ref{fig:coherent fraction no diffraction}b and Fig.~\ref{fig: theory experiment comparison}.

\subsection{Behavior of the coherent fraction for Gaussian correlations in various limits}

In this section, we consider the behavior of Eq.~\ref{eq: GS free coherent fraction} in various limits (returning to the case of zero time-delay $\tau = 0$). First, removing the aperture $a \to \infty$ with the width $\sigma$ and coherence length $\xi$ fixed yields
\begin{equation}
    C_{\text{GS}} = \frac{1}{1 + 2 \sigma^2 / \xi^2}
\end{equation}
which can be verified with the Fresnel propagation integral Eq.~\ref{eq: fresnel}. Using the Fraunhofer propagation integral Eq.~\ref{eq: fraunhofer}, the coherent fraction is $C_{\text{GS}} = \xi^2 / 2 \sigma^2$. This result matches the asymptotic behavior of Eq.~\ref{eq: GS free coherent fraction} for $\xi \ll \sigma$; the lack of an aperture limits the applicability of the Fraunhofer approximation. On the other hand, the limit of uniform intensity $\sigma \to \infty$ with the aperture radius $a$ and coherence length $\xi$ fixed yields 
\begin{equation}
    C_{\text{GS}} = \frac{\xi^2}{4 a^2} \left( 1 - e^{- 4 a^2 / \xi^2} \right)
\end{equation}
In contrast to the limit of no aperture, this result can be verified with the Fraunhofer propagation integral Eq.~\ref{eq: fraunhofer}. Lastly, the limit of complete coherence $\xi \to \infty$ with the width $\sigma$ and aperture radius $a$ fixed yields the expected result $C_{\text{GS}} = 1$.

\subsection{Behavior of the coherent fraction for Bose gas correlations in various limits} \label{sec: coherent fraction limits}

The behavior of Eq.~\ref{eq: BG free coherent fraction} for small fugacities $z \ll 1$ is easy to obtain since only a few terms in the sum are necessary. Therefore the limiting behavior roughly matches that of Eq.~\ref{eq: GS free coherent fraction} described above. For larger fugacities, in particular in the limit $z \to 1$, the calculation becomes cumbersome, and it is helpful to make an integral approximation. In this section of the appendix, we consider the behavior of Eq.~\ref{eq: BG free coherent fraction} at finite fugacity in the limit of no aperture and in the limit of uniform intensity.

\subsubsection{Limit of no aperture}

We start from the coherent fraction in Eq.~\ref{eq: BG free coherent fraction} and obtain
\begin{equation}
    C_{\text{BG}} = \frac{1}{- \log \left( 1 - z \right)} \sum_{j = 1}^\infty \frac{ z^j }{ j + 2 \sigma^2 / \xi_0^2 }
\end{equation}
in the limit of no aperture $a \to \infty$ with the width $\sigma$ and the classical coherent length $\xi_0$ fixed. Using the Euler-Maclaurin formula, which works well for $z \to 1$, we make the approximation
\begin{align} \label{eq: BG beam coherent fraction Euler-Maclaurin}
    C_{\text{BG}} & \approx \frac{1}{- \log \left( 1 - z \right)} \left( \int_1^\infty dj \frac{ z^j }{ j + 2 \sigma^2 / \xi_0^2 } + \frac{z}{2 + 4 \sigma^2 / \xi_0^2} \right) \\
    & \approx \frac{1}{- \log \left( 1 - z \right)} \left( \frac{ - \text{li} \left( z^{ 1 + 2 \sigma^2 / \xi_0^2 } \right) }{ z^{ 2 \sigma^2 / \xi_0^2 } } + \frac{z}{2 + 4 \sigma^2 / \xi_0^2} \right)
\end{align}
where $\text{li}$ is the logarithmic integral. Finally, we expand about $z = 1$ to obtain the limiting behavior
\begin{equation}
    C_{\text{BG}} \sim \frac{1}{z^{2 \sigma^2 / \xi_0^2}} \left( 1 - \frac{ \log \left( 1 + 2 \sigma^2 / \xi_0^2 \right) + \gamma }{ - \log \left( 1 - z \right) } \right)
\end{equation}
where $\gamma$ is the Euler-Mascheroni constant. This expression is accurate to the order $10^{-3}$. 

We can also investigate the behavior of the coherent fraction $C_{\text{BG}}$ with the width $\sigma$ and the classical coherence length $\xi_0$ at small but finite fugacity $0 < z \ll 1$ via a series expansion. First, in the limit $\xi_0 \gg \sigma$ we find
\begin{equation}
    C_{\text{BG}} = \frac{1}{- \log \left( 1 - z \right)} \sum_{j = 0}^\infty \left( -1 \right)^j 2^j \left( \frac{\sigma}{\xi_0} \right)^{2j} \text{Li}_{j + 1} \left(z \right)
\end{equation}
where $\text{Li}_j$ is the polylogarithm of order $j$. The leading-order term $j = 0$ ensures that $C_{\text{BG}} = 1$ in the limit $\xi_0 \to \infty$ with $\sigma$ fixed since $\text{Li}_1 \left( z \right) = - \log \left( 1 - z \right)$. Similarly, in the limit $\xi_0 \ll \sigma$ we find
\begin{equation}
    C_{\text{BG}} = \frac{1}{- \log \left( 1 - z \right)} \frac{1}{2} \left( \frac{\xi_0}{\sigma} \right)^2 \sum_{j = 0}^\infty \frac{\left( -1 \right)^j}{2^j} \left( \frac{\xi_0}{\sigma} \right)^{2j} \text{Li}_{-j} \left(z \right)
\end{equation}
The leading-order term $j = 0$ yields quadratic scaling which matches the Gaussian-Schell result $C_{\text{GS}}$ in the limit $z \to 0$.

\subsubsection{Limit of uniform intensity}

We start from the coherent fraction in Eq.~\ref{eq: BG free coherent fraction} and obtain 
\begin{equation}
    C_{\text{BG}} = \frac{1}{- \log \left( 1 - z \right)} \frac{\xi_0^2}{4 a^2} \sum_{j = 1}^\infty z^j \left( 1 - e^{- 4 a^2 / j \xi_0^2} \right)
\end{equation}
in the limit of uniform intensity $\sigma \to \infty$ with the aperture radius $a$ and the classical coherence length $\xi_0$ fixed. We can evaluate the series in two parts. The first is a geometric series. For the second part, we use the Euler-Maclaurin formula which works well in the limit $z \to 1$
\begin{equation}
\begin{split}
    C_{\text{BG}}  & \approx \frac{1}{- \log \left( 1 - z \right)} \frac{\xi_0^2}{4 a^2} \left( \frac{z}{1 - z} - \int_1^\infty dj z^j e^{- 4 a^2 / j \xi_0^2} - \frac{1}{2} e^{- 4 a^2 / \xi_0^2} \right)
\end{split}
\end{equation}
To evaluate the integral, we extend the lower integration bound from $1$ to $0$ since the integrand is negligible for $j \ll 4 a^2 / \xi_0^2$ and usually $4 a^2 / \xi_0^2 \gg 1$ for experimental systems. Then
\begin{equation}
    \int_0^\infty dj z^j e^{- 4 a^2 / j \xi_0^2} = \frac{ 4 a K_1 \left( 4 a \sqrt{ - \log \left( z \right) } / \xi_0 \right) }{ \xi_0 \sqrt{ - \log \left( z \right) }}
\end{equation}
where $K_1$ is the first-order modified Bessel function of the second kind. Then
\begin{equation}
    C_{\text{BG}} \approx \frac{1}{- \log \left( 1 - z \right)} \frac{\xi_0^2}{4 a^2} \left( \frac{z}{1-z} - \frac{ 4 a K_1 \left( 4 a \sqrt{ - \log \left( z \right) } / \xi_0 \right) }{ \xi_0 \sqrt{ - \log \left( z \right) } } \right)
\end{equation}
also neglecting the boundary term from the Euler-Maclaurin formula. Finally, we expand about $z = 1$ to obtain the limiting behavior
\begin{equation} \label{eq: BG coherent fraction limiting behavior}
    C_{\text{BG}} \sim 1 - \frac{ 2 \log \left( 2 a / \xi_0 \right) }{ - \log \left( 1 - z \right) } 
\end{equation}
which is accurate to the order of $10^{-3}$. 

We can also investigate the behavior of the coherent fraction $C_{\text{BG}}$ with the aperture radius $a$ and the classical coherent length $\xi_0$ at small but finite fugacity $0 < z \ll 1$ via a series expansion. First, in the limit $\xi_0 \gg a$ we expand the exponential in a power series and simplify the sum to obtain
\begin{equation}
    C_{\text{BG}} = \frac{1}{- \log \left( 1 - z \right) } \sum_{j = 0}^\infty \frac{ \left( - 1 \right)^j }{  \left( j + 1 \right)! } \left( \frac{ 2 a}{ \xi_0} \right)^{2 j} \text{Li}_{ j + 1 } \left( z \right)
\end{equation}
The leading-order term $j = 1$ ensures that $C_{\text{BG}} = 1$ in the limit $\xi_0 \to \infty$ with $a$ fixed since $\text{Li}_1 \left( z \right) = - \log \left( 1 - z \right)$. Second, in the limit $\xi_0 \ll a$, we ignore the exponential term $e^{- 4 a^2 / j \xi_0^2}$ with the assumption that the fugacity is sufficiently small such that $z^j$ suppresses the exponential term once the summation index $j$ is on the order $j \sim \xi_0^2 / 4 a^2$. Then 
\begin{equation}
    C_{\text{BG}} \approx \frac{1}{- \log \left( 1 - z \right)} \frac{z}{1 - z} \frac{ \xi_0^2}{4 a^2}
\end{equation}
The limiting value for $z \to 0$ matches the limiting value of the Gaussian-Schell result $C_{\text{GS}}$.

\subsection{Coherent fraction at finite time-delay} \label{sec: coherent fraction finite time-delay}

Following the procedure outlined in Sec.~\ref{sec: results}, we can calculate the coherent fraction as a function of the time-delay $\tau$ for a classical Maxwell-Boltzmann source $z \to 0$ over a disk of radius $a$ in the limit of uniform intensity
\begin{equation} \label{eq: GS free space-time coherent fraction}
    C_\text{GS} \left( \tau \right) = \frac{\xi^2}{4 a^2} \left( 1 - e^{ - 4 a^2 / \xi^2 \left( 1 + \tau^2 / \theta^2 \right) } \cos \left( \frac{ 4 \tau / \theta }{ \xi^2 \left( 1 + \tau^2 / \theta^2 \right) } \right) \right)
\end{equation}
Therefore, the coherent fraction $C_\text{GS}$ can exhibit oscillations in both the coherence length $\xi$ (with $\tau \neq 0$) and the time-delay $\tau$ (with $\xi$ finite). This is in contrast to the correlation area which is more well-behaved
\begin{equation}
    \int d^2 r \left| g^{(1)} \left( r, \tau \right) \right|^2 = \frac{\pi \xi^2}{2} \left( 1 - e^{ -  2 a^2 / \xi^2 \left( 1 + \tau^2 / \theta^2 \right) } \right)
\end{equation}
In the limit of no aperture $a \to \infty$ with the coherence length $\xi$ and time-separation $\tau$ fixed, the correlation area becomes $\pi \xi^2 / 2$ which is independent of the time-separation $\tau$ as the disk no longer limits which points can contribute to the correlations. Similarly, the coherence time is
\begin{equation}
    \int d \theta \left| g^{(1)} \left( r = 0, \tau \right) \right|^2 = \pi
\end{equation}
which is quoted in the main text.

The coherent fraction at finite time-separations in the case of Bose gas correlations can be constructed similarly to Eq.~\ref{eq: BG free coherent fraction}
\begin{equation} \label{eq: BG free space-time coherent fraction}
    C_\text{BG} \left( \tau \right) = \frac{1}{ - \log \left( 1 - z \right) } \sum_{j = 1}^\infty \frac{z^j}{j} C_\text{GS} \left( \tau, j \theta, \sqrt{j} \xi_0 \right)
\end{equation}
where $C_\text{GS} \left( \tau, j \theta, \sqrt{j} \xi_0 \right)$ denotes the classical Maxwell-Boltzmann result Eq.~\ref{eq: GS free space-time coherent fraction} with the adjusted thermal time-scale $j \theta$ and the adjusted coherence length $\sqrt{j} \xi_0$. The expression for $C_\text{BG} \left( \tau \right)$ does not yield oscillations in the fugacity $z$ or equivalently the density $n$, a fact which is already clear from Eq.~\ref{eq: space-time first-order coherence function}.

\section{Diffraction of partially coherent sources} \label{sec: diffraction of partially coherent sources}

In deriving Eq.~\ref{eq: GS free coherent fraction} and \ref{eq: BG free coherent fraction}, we assume that we can simply integrate over a disk of radius $a$. However, this disk is physically implemented as a circular aperture in the imaging system. While the precise diffraction pattern is not needed to calculate the coherent fraction, as noted in Sec.~\ref{sec: coherent fraction conservation}, we model the diffraction of the partially coherent source in order to gain insight into the nature of the experimental signal and investigate diffraction losses.

The intensity of partially coherent light after a physical (finite-sized) circular aperture has been investigated for various functional forms of correlations including Gaussian and Bessel-function \cite{shore_diffraction_1966, shore_effect_1968, singh_partially_1971}. Moreover, the mutual intensity function (equal-time first-order coherence function) has been investigated for two-slit interference \cite{schouten_new_2003, gbur_youngs_2022}. To the best of our knowledge, propagated Bose gas correlations have not been investigated. 

\subsection{Diffraction pattern of sources with Gaussian or Bose gas correlations} \label{sec: diffraction with Bose gas correlations}

\begin{figure*}[t]
    \centering
    \includegraphics[width=1\textwidth, center]{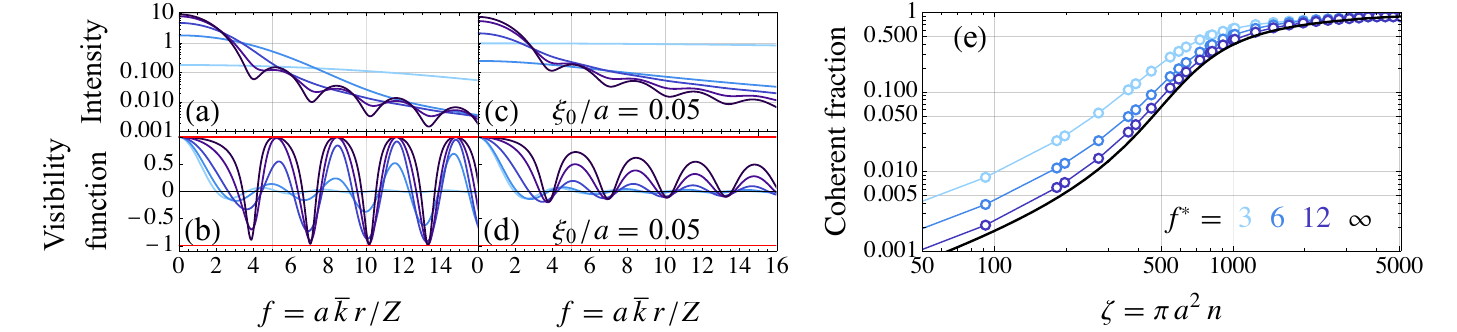}
    \caption{Analysis of a homogeneous, partially coherent source diffracted by a circular aperture. (a) Intensity of a source with Gaussian correlations in units of $\left( \epsilon_0 c / 2 \right) \left( \left| A \right|^2 a^4 / \bar{\lambda}^2 Z^2 \right)$ for various coherence lengths relative to the aperture radius $\xi / a = 0.14$, $0.5$, $1$, $2$, and $3.5$ (lighter blue for smaller $\xi / a$ and darker blue for larger $\xi / a$). (b) Visibility function for the same source and coherence lengths as (a) showing oscillations between regions of correlation and anti-correlation. The visibility function is bounded by $\pm 1$ (solid red). 
    (c) Intensity of a source with Bose gas correlations in units of $\left( \epsilon_0 c / 2 \right) \left( \left| A \right|^2 a^4 / \bar{\lambda}^2 Z^2 \right)$ for various particle numbers in the aperture $\zeta = \pi a^2 n = 10$, $1540$, $3185$, $5890$, and $11125$ (lighter blue for smaller $\zeta$ and darker blue for larger $\zeta$), equivalently treated as fugacities $z = 1 - \exp \left( - \xi_0^2 \zeta / a^2 \right)$ with the classical coherence length relative to the aperture radius $\xi_0 / a = 0.05$ fixed. (d) Visibility function for the same source and particle numbers in the aperture as (c) showing oscillations between regions of correlation and weak anti-correlation. The visibility function is bounded by $\pm 1$ (solid red). 
    (e) Coherent fraction of a source with Bose gas correlations with diffraction losses (solid blue and markers) calculated from the intensity and visibility function in (c) and (d), respectively. Integration is performed to $f^* = a \bar{k} r^* / Z$ where $r^*$ is the position in the imaging plane, and the coherent fraction for $\zeta > 0$ is taken relative to the zero density $\zeta = 0$ limit which corresponds to the classical Gaussian result. The results with diffraction losses converge to the exact coherent fraction (solid black) in the limit $f^* \to \infty$.
    }
    \label{fig:optics and diffraction}
\end{figure*}

We define the (real) transmission function of the aperture as the indicator function of the aperture
\begin{equation} \label{eq: aperture}
    P \left( \bm{s} \right) = \Theta \left( a - s \right) = \left\{ \begin{matrix} 1 & s \leq a \\ 0 & s > a \end{matrix} \right. 
\end{equation}
where $\Theta$ is the Heaviside step-function and $a$ is the aperture radius. If $a$ is sufficiently small relative to the beam width of the incident optical field, then the field can be approximated as having a uniform intensity over the aperture, i.e. taking the limit $\sigma / a \to \infty$. Propagating the initial mutual intensity function in the Fraunhofer limit with Eq.~\ref{eq: fraunhofer} assuming the limit of uniform intensity, we obtain the mutual intensity function of the diffracted light in the far-field from which we can extract the intensity and visibility function. The simplified propagation integrals are presented in Sec.~\ref{sec: propagation integrals}.

In Fig.~\ref{fig:optics and diffraction}a and b, we plot the intensity and visibility function in the far-field for a source with Gaussian correlations ($G^{(1)}_0$ leading to Eq.~\ref{eq: GS free coherent fraction} in the main text)
\begin{equation}
    \Gamma_0 \left( \bm{s}_1, \bm{s}_2 \right) = \left| A \right|^2 e^{ - \left( \bm{s}_1 - \bm{s}_2 \right)^2 / \xi^2}
\end{equation}
incident on an aperture, where $A$ is the (complex) amplitude of the scalar field and $\xi$ is the coherence length. The visibility function is always unity on-axis ($f = 0$). More generally, Ref.~\cite{schouten_new_2003} argues that equidistant points in Young's double slit experiment exhibit complete coherence regardless of the state of coherence of the illuminating light. Conversely, there are points off-axis ($f > 0$) which exhibit complete incoherence $V = 0$ regardless of the initial state of coherence of the optical field (so long as it is not completely coherent). In between, there are oscillations, known to occur in Young's double slit experiment \cite{schouten_phase_2003}, which we stress are physically distinct from the fringes observed in interferometry experiments. While fringes arise from the geometry of the interferometer (careful misalignment of interfering beams), the oscillations in the visibility are a result of the partial coherence of the illuminating source. In experiment, this suggests amplitude modulation of fringes. However, if measurements take place in the very far-field $Z \gg a \bar{k} r$ and the intensity becomes too low away from the axis, then only the near-axis region is considered.

A similar optical analysis can be performed for a source with Bose gas correlations ($G_0^{(1)}$ leading to Eq.~\ref{eq: BG free coherent fraction} in the main text)
\begin{equation}
    \Gamma_0 \left( \bm{s}_1, \bm{s}_2 \right) = \frac{ \left| A \right|^2 }{ - \log \left(1 - z \right) } \sum_{j = 1}^\infty \frac{ z^j }{j} e^{ - \left( \bm{s}_1 - \bm{s}_2 \right)^2 / j \xi_0^2} 
\end{equation}
incident on an aperture. The intensity is plotted in Fig.~\ref{fig:optics and diffraction}c while the visibility function is plotted in Fig.~\ref{fig:optics and diffraction}d, exhibiting similar oscillations. A qualitative difference, however, is that the strength of the anti-correlations (negative values) are much weaker. In contrast, Fig.~\ref{fig:optics and diffraction}b shows oscillations between correlations and anti-correlations which are roughly equal in magnitude.

In Ref.~\cite{alnatah_coherence_2024}, the diffraction pattern is focused by a lens which imposes a second circular aperture on the signal; the coherent fraction is then measured in the focal plane where the optical field is the Fourier transform of the source, assuming an infinite lens pupil. Relaxing this assumption by introducing a finite lens radius, but ensuring it is large enough to limit significant diffraction effects,
we study the corrections to the coherent fraction by numerically integrating the intensities and visibility functions from Fig.~\ref{fig:optics and diffraction}c-d. The result is shown in Fig.~\ref{fig:optics and diffraction}e for various integration limits. 
In general, a smaller integration window increases the measured coherent fraction since the mutual intensity decays more quickly than the intensity.

\subsection{Simplified propagation integrals} \label{sec: propagation integrals} 

In this section, we discuss in detail the propagation of an initial mutual intensity function past a circular aperture with transmission function Eq.~\ref{eq: aperture}. As stated before, we will assume the Fraunhofer limit and use Eq.~\ref{eq: fraunhofer}.

First, consider the case of Gaussian correlations. The four-dimensional propagation integral for the mutual intensity function in the Fraunhofer limit Eq.~\ref{eq: fraunhofer} is reduced to a one dimensional integral in the case of the intensity and a two dimensional integral in the case of the mutual intensity function evaluated at the points $\bm{r}_1 = - \bm{r}_2 = \bm{r}$. We list the simplified propagation integrals here. 

The intensity is
\begin{equation}
    I \left( \bm{f} \right) = \frac{\epsilon_0 c}{2} \frac{ 2 \pi^{3/2} \left| A \right|^2 a^4}{ \bar{\lambda}^2 Z^2 } \int_0^2 d\rho \; \rho e^{- \rho^2 / \xi_r^2 } G^{0 2}_{0 2} \left( \begin{matrix} 1/2 & 1 \\ -1 & 0 \end{matrix} \; \right| \left. \frac{4}{\rho^2} \right) J_0 \left( f \rho \right)
\end{equation}
where $G$ is the Meijer $G$-function, $J_0$ is the zeroth-order Bessel function of the first kind, $\xi_r = \xi / a$ is the coherence length relative to the aperture radius, and $\bm{f} = a \bar{k} \bm{r} / Z$ is the dimensionless far-field position with magnitude $f$. This expression is derived by applying Schell's theorem \cite{schell_multiple_1961, nugent_generalization_1990}. There does exist an exact series expansion for the intensity \cite{helstrom_detection_1969}, however the integral expression was numerically faster and more accurate in our studies. 

The mutual intensity function is derived from the convolution theorem. The general result of applying the convolution theorem to Eq.~\ref{eq: fraunhofer} and changing variables is
\begin{equation} \label{eq: convolution}
    \Gamma \left( \bar{\bm{f}}, \delta \bm{f} \right) = \frac{ \left| A \right|^2 a^2}{\bar{\lambda}^2 Z^2 n} e^{i \frac{Z}{\bar{k}} \bar{\bm{f}} \cdot \delta \bm{f}} \int d^2 \bar{g} \: \tilde{n} \left( \bar{\bm{f}} - \bar{\bm{g}} \right) \left( \frac{ J_1 \left( \left| \delta \bm{f} / 2 + \bm{g} \right| \right) }{ \left| \delta \bm{f} / 2 + \bm{g} \right| } \right) \left( \frac{ J_1 \left( \left| \delta \bm{f} / 2 - \bm{g} \right| \right) }{ \left| \delta \bm{f} / 2 - \bm{g} \right| } \right)
\end{equation}
where $\tilde{n}$ is the momentum distribution, as discussed in the main text, and we have normalized by the (homogeneous) density $n$. The mutual intensity function is best cast in terms of the mean $\bar{\bm{f}} = \left( \bm{f}_1 + \bm{f}_2 \right) / 2$ and difference $\delta \bm{f} = \bm{f}_1 - \bm{f}_2$ of the dimensionless far-field position variables $\bm{f}_1 = a \bar{k} \bm{r}_1 / Z$ and $\bm{f}_2 = a \bar{k} \bm{r}_2 / Z$. In the classical Maxwell-Boltzmann limit, we use the Gaussian momentum density $\tilde{n} \left( \bm{k}_\parallel \right) = e^{- \xi_r^2 k^2_\parallel / 4}$ and the density $n = 1 / \lambda_T^2$. Evaluating at $\bm{r}_1 = - \bm{r}_2 = \bm{r}$ or $\bar{\bm{f}} = 0$ we obtain
\begin{equation} \label{eq: gaussian propagation}
\begin{split}
    \Gamma \left( \bm{0}, 2 \bm{f} \right) & = \frac{ \pi \left| A \right|^2 a^4 \xi_r^2}{ \bar{\lambda}^2 Z^2 } \int_0^\infty dg \int_0^{2 \pi} d\eta \; g e^{- \xi_r^2 g^2 / 4} \left( \frac{ J_1 \left( \left( f^2 + g^2 + 2 f g \cos \left( \eta \right) \right)^{1/2} \right) }{ \left( f^2 + g^2 + 2 f g \cos \left( \eta \right) \right)^{1/2} } \right) \\
    & \hspace{3cm} \cdots \times \left( \frac{ J_1 \left( \left( f^2 + g^2 - 2 f g \cos \left( \eta \right) \right)^{1/2} \right) }{ \left( f^2 + g^2 - 2 f g \cos \left( \eta \right) \right)^{1/2} } \right)
\end{split}
\end{equation}
where $\eta$ is geometrically the angle between $\bm{f}$ and $\bm{g}$. Although the expressions look different, the intensity and mutual intensity function are equal on-axis $\bm{r} = 0$ with the value
\begin{equation}
    I \left( \bm{0} \right) = \Gamma \left( \bm{0}, \bm{0} \right) = \frac{ \pi^2 \left| A \right|^2 a^4 \xi_r^2}{ \bar{\lambda}^2 Z^2 } \left( 1 - e^{- 2 / \xi_r^2} \left( I_1 \left( \frac{2}{\xi_r^2} \right) + I_0 \left( \frac{2}{\xi_r^2} \right) \right) \right)
\end{equation}
where $I_\nu$ is the $\nu$-order modified Bessel function of the first kind. 

Next, consider the case of Bose gas correlations. Since the exact mutual intensity function is expressed as a series of Gaussian functions, i.e. Eq.~\ref{eq: coherence function}, we could apply the Gaussian results to the problem at hand. However, this is computationally intense if the classical coherence length $\xi_0$ is small compared to the aperture radius $a$ as many nontrivial terms become important for $z \to 1$. Alternatively, if the classical coherence length $\xi_0$ is a sufficiently large fraction of the aperture radius $a$, then there exists $j \geq 1$ not too large such that the diffracted mutual intensity function may be approximated by the analytically known diffraction pattern for a completely coherent source. 

But since we are primarily interested in the former case, we instead directly use the Bose-Einstein momentum distribution in Eq.~\ref{eq: convolution}. This approach allows us to probe the parameter-space more quickly. We list the simplified propagation integrals here. The intensity is
\begin{equation} \label{eq: bose intensity propagation}
    I \left( \bm{f} \right) = \frac{\epsilon_0 c}{2} \frac{\pi \left| A \right|^2 a^4 \xi_{0, r}^2}{ \bar{\lambda}^2 Z^2 \left( - \log \left( 1 - z \right) \right) } \int_0^\infty dg \int_0^{2\pi} d \eta \: \frac{ g }{ z^{-1} e^{ \xi_{0, r}^2 \left( f^2 + g^2 - 2 f g \cos \left( \eta \right) \right) / 4 } - 1 } \left( \frac{ J_1 \left( g \right) }{ g } \right)^2
\end{equation}
where now $\xi_{0, r} = \xi_0 / a$ is the classical coherence length relative to the aperture radius. The mutual intensity function evaluated at $\bm{r}_1 = - \bm{r}_2 = \bm{r}$ or $\bar{\bm{f}} = 0$ is
\begin{equation} \label{eq: bose mutual propagation}
\begin{split}
    \Gamma \left( \bm{0}, 2 \bm{f} \right) & = \frac{\pi \left| A \right|^2 a^4 \xi_{0, r}^2}{ \bar{\lambda}^2 Z^2 \left( - \log \left( 1 - z \right) \right) } \int_0^\infty dg \int_0^{2\pi} d \eta \: \frac{ g }{ z^{-1} e^{ \xi_{0, r}^2  g^2 / 4 } - 1 } \left( \frac{ J_1 \left( \left( f^2 + g^2 + 2 f g \cos \left( \eta \right) \right)^{1/2} \right) }{ \left( f^2 + g^2 + 2 f g \cos \left( \eta \right) \right)^{1/2} } \right) \\
    & \hspace{3cm} \cdots \times \left( \frac{ J_1 \left( \left( f^2 + g^2 - 2 f g \cos \left( \eta \right) \right)^{1/2} \right) }{ \left( f^2 + g^2 - 2 f g \cos \left( \eta \right) \right)^{1/2} } \right)
\end{split}
\end{equation}

For comparison to experiment, it is more instructive to use the density rather than the fugacity as a parameter. Introducing the particle number in the aperture $\zeta = \pi a^2 n$, we can rewrite the integral expressions for the intensity and mutual intensity function in a non-dimensional form with the substitution $\zeta = - \log \left( 1 - z \right) / \xi_{0, r}^2$ and $z = 1 - e^{- \xi_{0, r}^2 \zeta}$. This is the parameterization used in the main text. With these expressions, the coherent fraction can be calculated numerically. 

\subsection{Correction to coherent fraction due to diffraction losses} \label{sec: coherent fraction finite window analytic}

As mentioned before, in Ref.~\cite{alnatah_coherence_2024} the optical signal from the EP propagates through an aperture and then a thin lens with the coherent fraction measurement made in the Fourier plane. We can estimate corrections to the coherent fraction due to losses at each optical element analytically. The propagation of a mutual coherence function $\Gamma_0$ through this imaging system is obtained using the Fresnel propagation integral Eq.~\ref{eq: fresnel}
\begin{equation}
\begin{split}
    \Gamma \left( \bm{q}_1, \bm{q}_2 \right) & = \frac{1}{ \bar{\lambda}^4 F^2 Z^2 } e^{i \frac{ \bar{k}}{2 F} \left( q_1^2 - q_2^2 \right) } \left\{ \int d^2 s_1 \: d^2 s_2 \: d^2 r_1 \: d^2 r_2 \sqrt{I \left( \bm{s}_1 \right) I \left( \bm{s}_2 \right) } g^{(1)} \left( \left| \bm{s}_1 - \bm{s}_2 \right| \right) P \left( \bm{s}_1 \right) P^* \left( \bm{s}_2 \right) \right. \\
    & \hspace{3cm} \left. \cdots \times e^{i \frac{ \bar{k}}{2 Z} \left( s_1^2 - s_2^2 \right) } P_\ell \left( \bm{r}_1 \right) P_\ell^* \left( \bm{r}_2 \right) e^{i \frac{ \bar{k}}{2 Z} \left( r_1^2 - r_2^2 \right) } e^{- i \frac{ \bar{k}}{Z} \left( \bm{r}_1 \cdot \bm{s}_1 - \bm{r}_2 \cdot \bm{s}_2 \right) } e^{- i \frac{ \bar{k}}{F} \left( \bm{q}_1 \cdot \bm{r}_1 - \bm{q}_2 \cdot \bm{r}_2 \right) } \right\}
\end{split}
\end{equation}
where $F$ is the focal length of the thin lens, $P_\ell$ is the pupil function of the thin lens, $\bm{q}_1$ and $\bm{q}_2$ are the position variables in the Fourier plane, position $Z + F$, $\bm{r}_1$ and $\bm{r}_2$ are the position variables at the thin lens, position $Z$, and finally $\bm{s}_1$ and $\bm{s}_2$ are the position variables in the aperture, position $0$, as defined in the previous section. If we assume no pupil function for the thin lens (i.e. an infinite thin lens), then we can perform the integrals of $\bm{r}_1$ and $\bm{r}_2$ to get 
\begin{equation}
    \Gamma \left( \bm{q}_1, \bm{q}_2 \right) = \frac{1}{ \bar{\lambda}^2 F^2 } e^{i \frac{ \bar{k}}{2 F} \left( 1 - \frac{Z}{F} \right) \left( q_1^2 - q_2^2 \right) } \left\{ \int d^2 s_1 \: d^2 s_2 \sqrt{I \left( \bm{s}_1 \right) I \left( \bm{s}_2 \right) } g^{(1)} \left( \left| \bm{s}_1 - \bm{s}_2 \right| \right) P \left( \bm{s}_1 \right) P^* \left( \bm{s}_2 \right) e^{ - i \frac{\bar{k}}{F} \left( \bm{q}_1 \cdot \bm{s}_1 - \bm{q}_2 \cdot \bm{s}_2 \right) }\right\}
\end{equation}
which is the standard result that a thin lens performs a Fourier transform of the incoming field by moving the focus at infinity to the focal plane. In this case, the coherent fraction is conserved, as shown in the previous section.

In experiment, however, the lens has a finite size which results in a partial loss of optical signal diffracted from the aperture and vignetting. Assuming that the lens is circular with a radius $a_\ell$ and neglecting finite size corrections due to the integration over the positions $\bm{q}_1$ and $\bm{q}_2$ in the Fourier plane, we can obtain corrections to the coherent fraction using Eq.~\ref{eq: coherent fraction propagated} with $W$ being the disk of radius $a_\ell$. In this case,
\begin{equation}
    \int_W d^2 r \; e^{- i \frac{\bar{k}}{Z} \bm{r} \cdot \left( \bm{s}_1 \pm \bm{s}_2 \right) }  = 2 \pi a_\ell^2 \frac{ J_1 \left( \bar{k} a_\ell \left| \bm{s}_1 \pm \bm{s}_2 \right| / Z \right) }{ \bar{k} a_\ell \left| \bm{s}_1 \pm \bm{s}_2 \right| / Z }
\end{equation}
where $J_\nu$ is the $\nu$-order Bessel function of the first kind. This result can be analyzed in the limit of an infinite lens $\bar{k} a_\ell \to \infty$ by expressing the first-order Bessel function as an integral 
\begin{equation}
    J_1 \left( \frac{ \bar{k} a_\ell \left| \bm{s}_1 \pm \bm{s}_2 \right| }{ Z } \right) = \frac{ \left| \bm{s}_1 \pm \bm{s}_2 \right| }{ \bar{k} a_\ell Z } \int_0^{\bar{k} a_\ell } dx \; x J_0 \left( \frac{ \left| \bm{s}_1 \pm \bm{s}_2 \right| }{ Z } x \right)
\end{equation}
and noting the Bessel function definition of the Dirac delta function
\begin{equation}
    \delta \left( y \right) = y \int_0^\infty dx \; x J_0 \left( y x \right) 
\end{equation}

\section{Additional theory-experiment comparison and connection to Gross-Pitaevskii simulation} \label{sec: additional data and GPE}

\subsection{Additional data}

\begin{figure}
    \centering
    \includegraphics[width=0.5\linewidth]{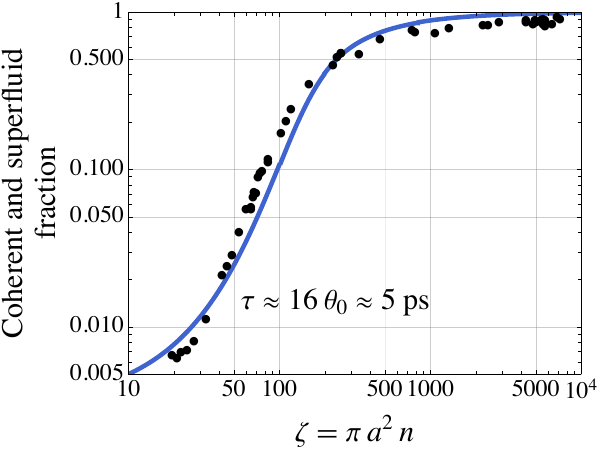}
    \caption{Comparison of the noninteracting, equilibrium theory presented in the main text to experimental results from Ref.~\cite{alnatah_coherence_2024} for a smaller aperture size than the data presented in the main text. The experimental coherent fraction measurements (black circles) are plotted in terms of the EP number in the aperture $\zeta = \pi a^2 n$.
    The theory prediction for the coherent fraction measured by an interferometer with a finite time-delay of approximately $5$~ps (solid dark blue) as given by Eq.~\ref{eq: BG free space-time coherent fraction} and its extensions is plotted at temperature $T = 24$~K. The result for finite time-delay becomes coincident with the $\tau = 0$ result at high densities as the coherence time becomes comparable to the time-delay. 
    The plot assumes an EP mass $m = 1.515 \times 10^{-4} m_e$, where $m_e = 9.109 \times 10^{-31}$~kg is the vacuum electron mass, and aperture radius of $a = 3.5$~$\mu$m. 
    }
    \label{fig: extended data}
\end{figure}

We also supply a comparison of the noninteracting, equilibrium theory presented in the main text to recent experimental result from Ref.~\cite{alnatah_coherence_2024} for different experimental conditions, in particular an aperture radius reduced from $a = 6$~$\mu$m to $3.5$~$\mu$m. The result is shown in Fig.~\ref{fig: extended data}. In theory, the reduced aperture radius should result in increased coherent fraction in the zero-density limit, i.e. the ``pinhole effect''. Since the results only depend on the dimensionless classical coherence length as $\xi_0 / a$, this is also equivalent to decreasing the EP temperature (strictly $\sqrt{T}$). For small pinhole sizes--corresponding to this case--the coherent fraction at low density is highly sensitive to the relative time-delay between the Michelson arms. We obtain the best fit when a delay of approximately $\tau \approx 5$~ps is introduced, which is consistent with the experimental uncertainty in the time-delay, as discussed in the main text.

\subsection{Connection to GPE simulation}

Previous studies $\cite{alnatah_coherence_2024}$ have used the Gross-Pitaevskii equation to model the build-up of the coherent fraction. The advantage of this approach is that it can interpolate between the low-density limit and the superfluid regime where interactions are important; however, the disadvantage is that it does not produce simple analytic results. 

Using the Gross-Pitaevskii equation, the mean-field EP wavefunction $\psi$ is initialized as an equilibrium state with random phases written as
\begin{equation}
\psi \left( \bm{r} \right) = \frac{1}{\sqrt{A}} \sum_{\bm{k}_\parallel} \sqrt{ \tilde{n} \left( \bm{k}_\parallel \right) } e^{i \bm{k}_\parallel \cdot \bm{r} }e^{i \theta_{\bm{k}_\parallel}}
\end{equation}
where $\tilde{n}$ is the Bose-Einstein momentum distribution, as in the main text, and $\theta_{\bm{k}_\parallel}$ are the random phases selected at each momenta. The initial wavefunction is then evolved in time, averaging over the random phases $\theta_{\bm{k}_\parallel}$. The coherent fraction is lastly extracted as
\begin{equation}
    C  = \frac{\int \mathrm{d}^2 r \; \psi^{*} \left( \bm{r} \right) \psi \left( - \bm{r} \right) }{\int \mathrm{d}^2 r \; \left| \psi \left( \bm{r} \right) \right|^2 }
\end{equation}
similar to Eq.~\ref{eq: coherent fraction} in the main text, but instead using the mean-field wavefunction rather than correlation functions. 

\end{document}